\documentclass[aps,prd]{revtex4}
\usepackage[dvips]{graphicx}
\usepackage[T1]{fontenc}
\usepackage[latin1]{inputenc}
\usepackage{amsmath}
\usepackage{amssymb}
\usepackage{mathrsfs}
\newcommand{\ud}{\mathrm{d}}
\newcommand{\dirac}{\partial\llap{$\diagup$\kern-2pt}}
\newcommand{\Dirac}{\mathscr{D}\llap{$\diagup$\kern-2pt}}
\newcommand{\fett}[1]{\boldsymbol{#1}}
\newcommand{\fettu}[1]{\mathbf{#1}}

\newcommand{\newatop}[3]{\raisebox{#3ex}{$\genfrac{}{}{0pt}{1}{#1}{#2}$}}

\newcommand{\diag}{\mathrm{diag}}

\newcommand{\Sp}{\mathrm{Tr}}
\renewcommand{\sp}{\mathrm{Tr}}

\begin{document}

\begin{abstract}
We compare quark stars made of color-superconducting
quark matter to normal-conducting quark stars.
We focus on the most simple color-superconducting system, a
two-flavor color superconductor, and employ the Nambu--Jona-Lasinio (NJL)
model to compute the gap parameter and the equation of state.
By varying the strength of the four-fermion coupling of the NJL model,
we study the mass and the radius
of the quark star as a function of the value of the gap parameter.
If the coupling constant exceeds a critical value, the gap parameter does
not vanish even at zero density. For coupling constants below this
critical value, mass and radius of a color-superconducting quark star
change at most by 
$\sim 20\%$ compared to a star consisting of normal-conducting
quark matter. For coupling constants above the critical value mass and
radius may change by factors of two or more.
\end{abstract}

\title{Effect of color superconductivity on the mass and radius of a quark star}

\author{Stefan B.\ R\"uster}
\email{ruester@th.physik.uni-frankfurt.de}
\affiliation{
Institut f\"ur Theoretische Physik, J.W.\ Goethe-Universit\"at,\\
D-60054 Frankfurt am Main, Germany}

\author{Dirk H.\ Rischke}
\email{drischke@th.physik.uni-frankfurt.de}
\affiliation{
Institut f\"ur Theoretische Physik, J.W.\ Goethe-Universit\"at,\\
D-60054 Frankfurt am Main, Germany}

\date{\today}

\maketitle

\section{introduction}
At sufficiently high densities and sufficiently low temperatures quark matter is a color superconductor \cite{BailinLove}.
In nature, color-superconducting quark matter could exist in the interior of compact stellar objects such as neutron or quark stars. Among the best known properties of compact stellar objects are their masses and radii. The question then is whether these observable properties allow to decide if compact stellar objects contain, or are even completely made of, color-superconducting quark matter. To this end, one has to compute these properties for stars containing color-superconducting quark matter and compare them to the corresponding ones for stars containing normal-conducting quark matter.

This question has recently triggered a lot of activity \cite{Prakash,Baldo,Steiner,buballaoertel,LugonesHorvath1,LugonesHorvath2,AlfordReddy,Grigorian1,Banik,Grigorian2,Blaschke,Shovkovy}.
Pure quark stars as well as hybrid stars were considered, both with two-flavor color-superconducting quark matter as well as with quark matter consisting of three flavors in the color-flavor-locked phase \cite{CFL}.
All these investigations are based on variants of the NJL model \cite{NJL} where fermions interact via a four-point vertex. The coupling strength is adjusted to be in agreement with hadron phenomenology at zero quark-chemical potential $\mu$. This leads to color-superconducting gap parameters $\phi$ of the order of 100 MeV \cite{CFL,RappSchaeferShuryakVelkovsky}. For reasonable values of the parameters entering the equation of state, such as the MIT bag constant $B$ and the strange quark mass $m_s$, the result of these studies is that mass and radius of a compact stellar object change by $\sim$ 20\%, if it contains color-superconducting instead of normal-conducting quark matter.

In this paper, we consider a different question. We ask \emph{how large} the color-superconducting gap parameter has to be in order to see \emph{substantial} changes in mass and radius of a compact stellar object. As the transition to hadronic matter introduces another degree of freedom which may either mask \cite{Prakash} or enhance \cite{AlfordReddy} the effects of color superconductivity, we do not consider hybrid stars, but focus exclusively on pure quark stars. We also consider the most simple color-superconducting state, namely quark matter with two flavors in the so-called 2SC phase, although this state may not be the most favorable one \cite{Absence}.

This paper is organized in the following way.
In Sec.\ \ref{eos} we derive the gap equation and the equation of state for two-flavor color-superconducting quark matter using the Cornwall-Jackiw-Tomboulis (CJT) formalism \cite{CJT}.
While this formalism is equivalent to other approaches to derive the gap equation and the equation of state, it is nevertheless
the most elegant way. Moreover, it also provides a general framework that allows one to go beyond the standard mean-field approximation (although this direction is not pursued in this work). In addition, it accounts for the possibility of
non-vanishing gluon background fields generated by condensation of quark Cooper-pairs. Our derivation presented in Sec.\ \ref{eos} puts special emphasis on this point, which has previously been neglected in the the derivation of the gap equation. In Sec.\ \ref{results} we compute the masses and radii of quark stars via the Tolman-Oppenheimer-Volkoff (TOV) equation. Section \ref{conclusions} concludes this paper with a summary of our results.

Our units are
$\hbar=c=k_B=1$. The metric tensor is
$g_{\mu\nu}=\diag\left(1,-1,-1,-1\right)$.
Four-vectors are denoted as $K^\mu=(k^ 0,\fettu{k})$, where $\fettu{k}$ is a three-vector with modulus $k=|\fettu{k}|$ and direction $\hat{\fettu{k}}=\fettu{k}/k$. We work in the imaginary-time formalism, i.e., the space-time integration is defined as $\int_X=\int_0^{1/T}\ud\tau\int_V\ud^3\fettu{x}$, where $\tau$ is Euclidean time, $T$ is the temperature, and $V$ the three-volume of the system. Energy-momentum sums are written as $T/V\sum_K=T\sum_n\int\ud^3\fettu{k}/(2\pi)^3$, where the sum runs over the Matsubara frequencies $\omega_n=2n\pi T$ for bosons and $\omega_n=(2n+1)\pi T$ for fermions, respectively.
\section{Equation of State and Gap Equation}
\label{eos}
For color-superconducting matter in the 2SC phase, the Lagrangian is given by
\begin{equation}
\label{2SC-L}
\mathscr{L}=-\,\frac14\,G_{\mu\nu}^aG_a^{\mu\nu}+
\bar\psi\left(i\Dirac-\hat{m}\right)\psi\; ,
\end{equation}
where 
\begin{equation}
G_{\mu\nu}^a=\partial_\mu A_\nu^a-\partial_\nu A_\mu^a+gf^{abc}A_\mu^b A_\nu^c
\end{equation}
is the gluon field strength tensor; $A_\mu^a$ is the vector potential for the gluon field, $f^{abc}$ are the structure constants of $SU(3)_c$, and $g$ is the strong coupling constant.
The fermion fields $\psi$ are $(4N_cN_f=24)$-dimensional spinors. Suppressing the Dirac structure we choose the following basis in color-flavor space:
\begin{equation}
  \label{spinorbasis}
  \psi=\left(
  \begin{array}{c}
     \psi_r^u \\
     \psi_r^d \\
     \psi_g^u \\
     \psi_g^d \\
     \psi_b^u \\
     \psi_b^d
  \end{array}
  \right)\, .
\end{equation}
The Dirac conjugate spinor is defined as $\bar\psi=\psi^\dagger\gamma_0$.
The covariant derivative is given by
$\mathscr{D}_\mu=\partial_\mu-igA_\mu^a T_a$, where $T_a$ are the generators of $SU(3)_c$, suitably generalized to our 6-dimensional color-flavor basis (\ref{spinorbasis}).
The quark mass matrix is $\hat{m}=\diag(m_u,m_d,m_u,m_d,m_u,m_d)$.
When computing quark star properties in Sec.\ \ref{results}, we also include electrons in order to achieve electrical neutrality. In some cases we also add non-interacting strange quarks.

In the treatment of superconducting systems it is advantageous to double the fermionic degrees of freedom by introducing Nambu-Gor'kov spinors
\begin{equation}
  \label{Psi}
  \bar\Psi=\left(\bar\psi,\bar\psi_C\right)\;,\qquad
  \Psi=\left(
  \begin{array}{c}
    \psi \\
    \psi_C
  \end{array}
  \right)\; ,
\end{equation}
where $\psi_C=C\bar\psi^T$ is the charge-conjugate spinor; $C$ is the charge-conjugation matrix. In this basis, the tree-level action can be written as
\begin{equation}
  I\left[\bar\Psi,\Psi,A\right]=-\,\frac14\int_X
  G_{\mu\nu}^a(X)G_a^{\mu\nu}(X)
  +\frac12\int_{X,Y}\bar\Psi(X)\,S_0^{-1}(X,Y)\,\Psi(Y)\;,
\end{equation}
where
\begin{equation}
  \label{S0hm1}
  S_0^{-1}(X,Y)=\left(\begin{array}{cc}
  i\Dirac_X+\hat\mu\gamma_0-\hat{m} & 0 \\
  0 & i\Dirac_X^{\;C}-\hat\mu\gamma_0-\hat{m}
  \end{array}\right)\delta^{(4)}(X-Y)
\end{equation}
is the tree-level propagator for Nambu-Gor'kov fermions. Here we introduced the charge-conjugate covariant derivative $\mathscr{D}_\mu^{\;C}=\partial_\mu+igA_\mu^a T_a^T$. The delta function is defined as $\delta^{(4)}(X-Y)=\delta(\tau_x-\tau_y)\,
\delta^{(3)}(\fettu{x}-\fettu{y})$.
The quark-chemical potential matrix is $\hat\mu=\diag(\mu_r^u,\mu_r^d,\mu_g^u,\mu_g^d,\mu_b^u, \mu_b^d)$. The chemical potential for quarks of color $i$ and flavor $f$ can be represented as
\begin{equation}
  \label{chemischePotentiale}
  \mu_i^f=\mu-\mu_e\,Q^f+\mu_3\,T_{ii}^3+\mu_8\,T_{ii}^8\;,
\end{equation}
where $\mu_e$ is the electro-chemical potential, $Q^f$ is the electric charge (in units of $e$) of quark flavor $f$, and $\mu_3$ and $\mu_8$ are the color-chemical potentials associated with the diagonal generators $T^3$ and $T^8$ of $SU(3)_c$.
While $\mu$ controls the quark number density, $\mu_e$, $\mu_3$, and $\mu_8$ have to be introduced to ensure electric- and color-charge neutrality. If the $SU(3)_c$ color symmetry is not broken, the color-chemical potentials have to vanish, $\mu_3=\mu_8=0$, otherwise they would break $SU(3)_c$ explicitly. However, when the color symmetry is broken by a color-charged quark Cooper-pair condensate, $\mu_3$ and $\mu_8$ do not need to be zero. We shall come back to this issue below.

\begin{figure}[!hbp]
  \begin{center}
    \includegraphics[scale=0.5]{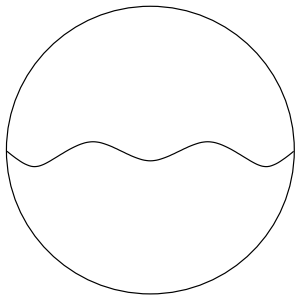}
    \caption{The sunset-type diagram.}
    \label{sunset}
  \end{center}
\end{figure}

The effective action in the CJT formalism \cite{CJT} reads \cite{MiranskiShovkovy,Tagaki,RischkeReview,Japaner}
\begin{eqnarray}
  \Gamma\left[\bar\Psi,\Psi,A,S,D\right]
  &=&I\left[\bar\Psi,\Psi,A\right]
  -\frac12\Sp\ln D^{-1}
  -\frac12\Sp\left(D_0^{-1}D-1\right)\nonumber\\
  &+&\frac12\Sp\ln S^{-1}
  +\frac12\Sp\left(S_0^{-1}S-1\right)
  +\Gamma_2\left[\bar\Psi,\Psi,A,S,D\right]\; .
  \label{Gamma}
\end{eqnarray}
The quantities $D$ and $S$ are the full gluon and quark propagators, respectively. The inverse tree-level quark propagator $S_0^{-1}$ was introduced in Eq.\ (\ref{S0hm1}).
Correspondingly, $D_0^{-1}$ is the inverse tree-level gluon propagator. The traces run over space-time, Nambu-Gor'kov, color, flavor, and Dirac indices. The factor $1/2$ in front of the fermionic one-loop terms compensates the doubling of the degrees of freedom in the Nambu-Gor'kov basis. The functional $\Gamma_2$ is the sum of all two-particle irreducible (2PI) diagrams. It is impossible to evaluate all 2PI diagrams exactly. However, the advantage of the CJT effective action (\ref{Gamma}) is that truncating the sum $\Gamma_2$ after a finite number of terms still provides a well-defined many-body approximation. Here we only include the sunset-type diagram shown in Fig.\ \ref{sunset},
\begin{equation}
  \label{Gamma2}
  \Gamma_2=-\,\frac{g^2}{4}\int_{X,Y}
  \newatop{\displaystyle\Sp}{NG,c,f,s}{-1}
  \left[\Gamma_a^\mu\,S(X,Y)\,\Gamma_b^\nu\,S(Y,X)\right]\,
  D_{\mu\nu}^{ab}(X,Y)\;,
\end{equation}
where the trace now runs only over Nambu-Gor'kov, color, flavor, and Dirac indices. The Nambu-Gor'kov vertices are defined as
\begin{equation}
  \Gamma_a^\mu=\left(\begin{array}{cc}
    \gamma^\mu T_a & 0 \\
    0 & -\gamma^\mu T_a^T
  \end{array}\right)\;.
\end{equation}
Later on, we shall approximate the gluon-exchange interaction between quarks by a point-like four-fermion coupling. This effectively removes dynamical gluon degrees of freedom, such that we do not need to worry about gauge fixing or possible ghost degrees of freedom. Therefore we already omitted the latter in Eq.\ (\ref{Gamma}).

The stationary points of the effective action (\ref{Gamma}) determine the expectation values of the one- and two-point functions,
\begin{equation}
  \label{stationary}
  \frac{\delta\Gamma}{\delta\bar\Psi}=0\;,\;\;
  \frac{\delta\Gamma}{\delta\Psi}=0\;,\;\;
  \frac{\delta\Gamma}{\delta A_\mu^a}=0\;,\;\;
  \frac{\delta\Gamma}{\delta D}=0\;,\;\;
  \frac{\delta\Gamma}{\delta S}=0\;.
\end{equation}
The first two equations yield the Dirac equation for the fermionic fields $\Psi$ and $\bar\Psi$ in the presence of the gluon field $A_\mu^a$. The solution is trivial, since fermionic, i.e.\ Grassmann-valued, fields do not have a (c-number) expectation value. The third equation is the Yang-Mills equation for the gluon field,
\begin{equation}
  \mathcal{D}_\nu^{ab}F^{\nu\mu}_b(X)
  =\frac{\delta}{\delta A_\mu^a(X)}\,
  \left[\frac12\,\Sp
  \left(D_0^{-1}D-S_0^{-1}S\right)-\Gamma_2\right]\;,
\end{equation}
where $\mathcal{D}_\nu^{ab}=\partial_\nu \delta^{ab}-g f^{abc} A_\nu^c(X)$ is the covariant derivative in the adjoint representation. The first two terms on the right-hand side are the contributions from gluon and fermion tadpoles \cite{Rebhan}. The functional derivative with respect to $A_\mu^a$ acting on the trace is nontrivial because of the dependence of the inverse tree-level propagators $D_0^{-1}$ and $S_0^{-1}$ on the gluon field, cf.\ Eq.\ (\ref{S0hm1}). The last term is non-zero if $\Gamma_2$ contains 2PI diagrams with an explicit dependence on $A_\mu^a$. It vanishes in our approximation (\ref{Gamma2}) for $\Gamma_2$. As shown in Ref.\ \cite{Rebhan} the solution of the Yang-Mills equation in the 2SC phase is a constant background field $A_\mu^a\sim g_{\mu 0}\delta^{a8}$. This background field acts like a color-chemical potential $\mu_8$ and provides the color-charge neutrality of the 2SC phase \cite{Rebhan}. 
Later on, we shall remove the gluon degrees of freedom by approximating the non-local gluon exchange with a point-like four-fermion coupling. The constant background field $A_\mu^a$ then disappears from the treatment, and the color-chemical potential $\mu_8$ assumes the role of the background field to ensure color neutrality.

The fourth equation (\ref{stationary}) is the Dyson-Schwinger equation for the gluon propagator,
\begin{equation}
  {D^{-1}}_{ab}^{\mu\nu}(X,Y)={D_0^{-1}}_{ab}^{\mu\nu}(X,Y)
  +\Pi_{ab}^{\mu\nu}(X,Y)\;,
\end{equation}
where
\begin{equation}
  \Pi_{ab}^{\mu\nu}(X,Y)=-2\,
  \frac{\delta\Gamma_2}{\delta D_{ba}^{\nu\mu}(Y,X)}
  =\frac{g^2}{2}\,
  \sp\left[\Gamma_a^\mu\,S(X,Y)\,\Gamma_b^\nu\,S(Y,X)\right]
\end{equation}
is the gluon self-energy. Since we shall approximate gluon exchange by a four-fermion coupling, we do not need to solve the Dyson-Schwinger equation for the gluon propagator.

The fifth equation (\ref{stationary}) is the Dyson-Schwinger equation for the quark propagator,
\begin{equation}
  \label{DSE}
  S^{-1}(X,Y)=S_0^{-1}(X,Y)+\Sigma(X,Y)\;,
\end{equation}
where
\begin{equation}
  \Sigma(X,Y)=
  2\,\frac{\delta\Gamma_2}{\delta S(Y,X)}=
  -g^2\,\Gamma_a^\mu\,S(X,Y)\,\Gamma_b^\nu\,
  D_{\mu\nu}^{ab}(Y,X)
\end{equation}
is the quark self-energy.
Assuming translational invariance, in momentum space the Dyson-Schwinger equation reads
\begin{equation}
  \label{DSEK}
  S^{-1}(K)=S_0^{-1}(K)+\Sigma(K)\;,\qquad
  \Sigma(K)=-\,g^2\,\frac{T}{V}\sum_Q
  \Gamma_a^\mu\,S(Q)\,\Gamma_b^\nu\,
  D_{\mu\nu}^{ab}(K-Q)\;.
\end{equation}
Let us introduce the Nambu-Gor'kov matrices
\begin{equation}
  \label{Sigma}
  S_0^{-1}=
    \left(\begin{array}{cc}
    \left[G_0^+\right]^{-1} & 0 \\
    0 & \left[G_0^-\right]^{-1}
  \end{array}\right)\;,\qquad\qquad
  \Sigma=
  \left(\begin{array}{cc}
    \Sigma^+ & \Phi^- \\
    \Phi^+ & \Sigma^-
  \end{array}\right)\;,
\end{equation}
where
\begin{subequations} 
\begin{eqnarray}
\left[G_0^+\right]^{-1}(K)&=&\gamma^\mu (K_\mu+gA_\mu^a T_a)+\hat\mu\,\gamma_0-\hat{m}\;,\\
\left[G_0^-\right]^{-1}(K)&=&\gamma^\mu (K_\mu-gA_\mu^a T_a^T)-\hat\mu\,\gamma_0-\hat{m}
\end{eqnarray}
\end{subequations}
are the inverse tree-level propagators for particles and charge-conjugate particles, respectively.
The quantities $\Sigma^\pm$ in Eq.\ (\ref{Sigma}) are the normal self-energies for particles and charge-conjugate particles, while $\Phi^\pm$ are the so-called anomalous self-energies.
The self-energies are related via $\Sigma^-(K)=C[\Sigma^+(-K)]^T C^{-1}$ and $\Phi^-(K)=\gamma_0[\Phi^+(K)]^\dagger\gamma_0$.
With the definitions (\ref{Sigma}), in Nambu-Gor'kov space the Dyson-Schwinger equation (\ref{DSEK}) has the solution
\begin{equation}
  \label{S}
  S=
  \left(\begin{array}{cc}
    G^+ & \Xi^- \\
    \Xi^+ & G^-
  \end{array}\right)\;,
\end{equation}
where
\begin{subequations}
\begin{eqnarray}
\label{G}
G^\pm&=&\left\{\left[G_0^\pm\right]^{-1}+\Sigma^\pm-\Phi^\mp \left(\left[G_0^\mp\right]^{-1}+\Sigma^\mp\right)^{-1} \Phi^\pm\right\}^{-1}\;,\\
\label{Xi}
\Xi^\pm&=&-\left(\left[G_0^\mp\right]^{-1}+\Sigma^\mp \right)^{-1}\Phi^\pm\,G^\pm\;.
\end{eqnarray}
\end{subequations}
Here $G^\pm$ are the propagators for quasiparticles and charge-conjugate quasiparticles, respectively, while $\Xi^\pm$ are the so-called anomalous propagators.

The gap equation for the color-superconducting gap parameter can be deduced from the $(21)$- or $(12)$-components of the Nambu-Gor'kov self-energy (\ref{Sigma}),
\begin{subequations}
\label{Gapgleichung}
\begin{eqnarray}
  \label{Phiplus}
  \Phi^+(K)&=&g^2\,\frac{T}{V}\sum_Q \gamma^\mu T_a^T\,
   \Xi^+(Q)\, \gamma^\nu T_b\, D_{\mu\nu}^{ab}(K-Q)\;,\\
  \label{Phiminus}
  \Phi^-(K)&=&g^2\,\frac{T}{V}\sum_Q \gamma^\mu T_a\, \Xi^-(Q)\,   \gamma^\nu T_b^T\, D_{\mu\nu}^{ab}(K-Q)\;.
\end{eqnarray}
\end{subequations}
It is sufficient to consider Eq.\ (\ref{Phiplus}), because Eq.\ 
(\ref{Phiminus}) follows from the relation $\Phi^-(K)=\gamma_0[\Phi^+(K)]^\dagger\gamma_0$.
In the color-flavor basis (\ref{spinorbasis}) the gap matrix in the 2SC phase reads
\begin{equation}
  \label{Phi}
    \Phi^\pm=
    \left(
     \begin{array}{cccc@{\hspace{0mm}\extracolsep{3mm}}cc}
       0 & 0 & 0 & \Delta_1^\pm & 0 & 0 \\
       0 & 0 & \Delta_2^\pm & 0 & 0 & 0 \\
       0 & \Delta_2^\pm & 0 & 0 & 0 & 0 \\
       \Delta_1^\pm & 0 & 0 & 0 & 0 & 0 \\
       0 & 0 & 0 & 0 & 0 & 0 \\
       0 & 0 & 0 & 0 & 0 & 0 
     \end{array}\right)\; .
\end{equation}
Here we have used the fact that red up-quarks form Cooper pairs with green down-quarks, with a (Dirac space) gap matrix $\Delta_1^\pm$, while green up-quarks form Cooper pairs with red down-quarks, with the gap matrix $\Delta_2^\pm$. A priori, $\Delta_1^\pm\neq\Delta_2^\pm$. As required by the overall antisymmetry of the spin-zero gap matrix \cite{BailinLove}, the right-hand side of Eq.\ (\ref{Phi}) is symmetric in color-flavor space.

The regular quark self-energy $\Sigma^\pm$ was computed in Ref.\ \cite{Manuel}. To leading order,
\begin{equation}
  \label{self-energy}
  \Sigma^+(K)=\Sigma^-(K)\simeq\frac{g^2}{9\pi^2}\,
  \gamma_0\,k_0\,\ln\left(\frac{g\mu}{|k_0|}\right)\;.
\end{equation}
This corresponds to a wave function renormalization factor in the quark propagator. In the QCD gap equation, it leads to subleading corrections which modify the prefactor of the color-superconducting gap parameter \cite{Rockefeller,Wang}. Since we ultimately do not consider the QCD gap equation, but the one in a simpler point-like four-fermion coupling model, we neglect the regular quark self-energy (\ref{self-energy}) in the following.

In order to proceed we compute the full inverse quark propagator (\ref{G}) with the gap matrix (\ref{Phi}), which is diagonal in the color-flavor basis (\ref{spinorbasis}),
\begin{eqnarray}
 \left[G^\pm\right]^{-1}&=&
 \diag\left(
  [{G_0^\pm}_r^u]^{-1}-\Delta_1^\mp\,
   {G_0^\mp}_g^d\,\Delta_1^\pm,\ 
  [{G_0^\pm}_r^d]^{-1}-\Delta_2^\mp\,
   {G_0^\mp}_g^u\,\Delta_2^\pm,\ 
  [{G_0^\pm}_g^u]^{-1}-\Delta_2^\mp\,
   {G_0^\mp}_r^d\,\Delta_2^\pm,\right. \nonumber\\
&&\hspace*{0.9cm}\left.[{G_0^\pm}_g^d]^{-1}-\Delta_1^\mp\,
   {G_0^\mp}_r^u\,\Delta_1^\pm,\ 
  [{G_0^\pm}_b^u]^{-1},\ 
  [{G_0^\pm}_b^d]^{-1}
  \right)\; ,
  \label{fullGinv}
\end{eqnarray}
where
\begin{equation}
  [{G_0^\pm}_i^f]^{-1}=\gamma^\mu K_\mu\pm \mu_i^f\gamma_0\;.
\end{equation}
At this stage, we have set the constant gluon background field $A_\mu^a=0$ (color neutrality can be achieved by adjusting the color-chemical potential $\mu_8$), and we have also neglected the small up- and down-quark masses.
The elements of the full inverse quark propagator (\ref{fullGinv}) have a simple physical interpretation. Consider, for instance, the red-up element $[{G^\pm}_r^u]^{-1}$. The presence of the color-superconducting condensate $\Delta_1^+$ (consisting of Cooper pairs of red up- and green down-quarks) modifies the propagation of red up-quarks, such that a red up-quark can be converted into a charge-conjugate green down-quark which continues to propagate and is then converted back into a red up-quark by the charge-conjugate condensate $\Delta_1^-$.

With Eq.\ (\ref{Phi}) and setting $\Sigma^\pm=0$, Eq.\ (\ref{Xi}) reads in the color-flavor basis (\ref{spinorbasis})
\begin{equation}
   \label{Ximatrix}
    \Xi^\pm=
    \left(
     \begin{array}{cccc@{\hspace{-1mm}\extracolsep{5mm}}cc}
       0 & 0 & 0 & {\Xi^\pm}_{rg}^{ud} & 0 & 0 \\
       0 & 0 & {\Xi^\pm}_{rg}^{du} & 0 & 0 & 0 \\
       0 & {\Xi^\pm}_{gr}^{ud} & 0 & 0 & 0 & 0 \\
       {\Xi^\pm}_{gr}^{du} & 0 & 0 & 0 & 0 & 0 \vspace*{1mm}\\
       0 & 0 & 0 & 0 & 0 & 0 \vspace*{1mm}\\
       0 & 0 & 0 & 0 & 0 & 0 
     \end{array}\right)\; ,
\end{equation}
where
\begin{equation}
  \label{Xielemente}
  {\Xi^\pm}_{rg}^{ud}=-\,{G_0^\mp}_r^u \Delta_1^\pm      {G^\pm}_g^d\;,\quad
  {\Xi^\pm}_{rg}^{du}=-\,{G_0^\mp}_r^d \Delta_2^\pm      {G^\pm}_g^u\;,\quad
  {\Xi^\pm}_{gr}^{ud}=-\,{G_0^\mp}_g^u \Delta_2^\pm      {G^\pm}_r^d\;,\quad
  {\Xi^\pm}_{gr}^{du}=-\,{G_0^\mp}_g^d \Delta_1^\pm      {G^\pm}_r^u\;.
\end{equation}
Neglecting effects from the breaking of $SU(3)_c$ due to Cooper pair condensation (these effects are of sub-subleading order in the QCD gap equation \cite{Rischkeselfenergy}),
the gluon propagator can be taken to be diagonal in adjoint color,
$D_{\mu\nu}^{ab}(K-Q)=\delta^{ab}D_{\mu\nu}(K-Q)$. Inserting this in the gap equation (\ref{Phiplus}) and performing the sum over adjoint colors we obtain the following four equations by identifying the non-trivial elements of the resulting color-flavor matrix:
\begin{subequations}
\begin{eqnarray}
  \Delta_1^+(K)&=&g^2\,\frac{T}{V}\sum_Q \gamma^\mu
   \left[\frac12\,{\Xi^+}_{gr}^{ud}(Q)
  -\frac16\,{\Xi^+}_{rg}^{ud}(Q)\right]\, \gamma^\nu
  D_{\mu\nu}(K-Q)\;,\\
    \Delta_2^+(K)&=&g^2\,\frac{T}{V}\sum_Q \gamma^\mu
   \left[\frac12\,{\Xi^+}_{gr}^{du}(Q)
  -\frac16\,{\Xi^+}_{rg}^{du}(Q)\right]\, \gamma^\nu
  D_{\mu\nu}(K-Q)\;,\\
  \Delta_2^+(K)&=&g^2\,\frac{T}{V}\sum_Q \gamma^\mu
   \left[\frac12\,{\Xi^+}_{rg}^{ud}(Q)
  -\frac16\,{\Xi^+}_{gr}^{ud}(Q)\right]\, \gamma^\nu
  D_{\mu\nu}(K-Q)\;,\\
  \Delta_1^+(K)&=&g^2\,\frac{T}{V}\sum_Q \gamma^\mu
   \left[\frac12\,{\Xi^+}_{rg}^{du}(Q)
  -\frac16\,{\Xi^+}_{gr}^{du}(Q)\right]\, \gamma^\nu
  D_{\mu\nu}(K-Q)\;.
\end{eqnarray}
\end{subequations}
In order to determine $\Delta_1^+$ and $\Delta_2^+$, only two of these four equations are necessary. These two equations can be combined to
\begin{subequations}
\label{Gapkombi}
\begin{eqnarray}
  \Delta_1^+(K)+3\,\Delta_2^+(K)
  &=&\frac43\,g^2\,\frac{T}{V}\sum_Q \gamma^\mu
   \,{\Xi^+}_{rg}^{ud}(Q)\,\gamma^\nu
  D_{\mu\nu}(K-Q)\;,\\
  3\,\Delta_1^+(K)+\Delta_2^+(K)
  &=&\frac43\,g^2\,\frac{T}{V}\sum_Q \gamma^\mu
   \,{\Xi^+}_{gr}^{ud}(Q)\,\gamma^\nu
  D_{\mu\nu}(K-Q)\;.
\end{eqnarray}
\end{subequations}
We now have to determine ${\Xi^+}_{rg}^{ud}$ and ${\Xi^+}_{gr}^{ud}$.
The Dirac structure of the color-superconducting gap matrices $\Delta_{1,2}^\pm$ is conveniently written in terms of energy-chirality projectors \cite{Rischke-Paper},
\begin{equation}
  \mathcal{P}_c^e(\fettu{k})=\frac14\,(1+c\gamma_5)\,
  (1+e\gamma_0\fett{\gamma}\cdot\hat{\fettu{k}})\;,
\end{equation}
where $c=\pm$ stands for the right/left-handed projection, and $e=\pm$ denotes the projection onto states of positive/negative energy. With these projectors the gap matrices can be written as \begin{equation}
  \label{Delta_n}
  \Delta_n^+(K)=\sum_{c,e}\phi_{nc}^{\phantom{n}e}(K)\,
  \mathcal{P}_c^e(\fettu{k})\;,\qquad
  \Delta_n^-(K)=\sum_{c,e}{\phi_{nc}^{\phantom{n}e}}^*(K)\,
  \mathcal{P}_{-c}^{-e}(\fettu{k})\;,\qquad n=1,2\;.
\end{equation}
With the definition of the quasiparticle energy for positive and negative energy states ($e=\pm$)
\begin{equation}
  \epsilon_{\fettu{k}}^e\left(\mu,\phi\right)
  =\sqrt{\left(k-e\mu\right)^2
  +\left|\phi\right|^2}\;,
  \label{epsilon}
\end{equation}
we can write the components of the full quark propagator of the quark colors participating in Cooper pairing as
\begin{subequations}
\label{Gplus}
\begin{eqnarray}
  {G^+}_r^u(K)&=&\sum_{c,e}
  \frac{\mathcal{P}_c^e(\fettu{k})}
  {(k_0+\delta\mu_1)^2
  -\left[\epsilon_\fettu{k}^e
  \left(\bar\mu,
  \phi_{1c}^{\phantom{1}e}\right)\right]^2}\;
  [{G_0^-}_g^d]^{-1}(K)\;,\\
  {G^+}_r^d(K)&=&\sum_{c,e}
  \frac{\mathcal{P}_c^e(\fettu{k})}
  {(k_0+\delta\mu_2)^2
  -\left[\epsilon_\fettu{k}^e
  \left(\bar\mu,
  \phi_{2c}^{\phantom{2}e}\right)\right]^2}\;
  [{G_0^-}_g^u]^{-1}(K)\;,\\
  {G^+}_g^u(K)&=&\sum_{c,e}
  \frac{\mathcal{P}_c^e(\fettu{k})}
  {(k_0-\delta\mu_2)^2
  -\left[\epsilon_\fettu{k}^e
  \left(\bar\mu,
  \phi_{2c}^{\phantom{2}e}\right)\right]^2}\;
  [{G_0^-}_r^d]^{-1}(K)\;,\\
  {G^+}_g^d(K)&=&\sum_{c,e}
  \frac{\mathcal{P}_c^e(\fettu{k})}
  {(k_0-\delta\mu_1)^2
  -\left[\epsilon_\fettu{k}^e
  \left(\bar\mu,
  \phi_{1c}^{\phantom{1}e}\right)\right]^2}\;
  [{G_0^-}_r^u]^{-1}(K)\;,
\end{eqnarray}
\end{subequations}
where
\begin{subequations}
\label{muquer}
\begin{eqnarray}
  \bar\mu&=&\frac{\mu_r^u+\mu_g^d}{2}=
  \frac{\mu_r^d+\mu_g^u}{2}=\mu-\frac{\mu_e}{6}
  +\frac{\mu_8}{2\sqrt{3}}\;,\\
  \delta\mu_1&=&\frac{\mu_r^u-\mu_g^d}{2}=
  \frac12\left(\mu_3-\mu_e\right)\;,\\
  \delta\mu_2&=&\frac{\mu_r^d-\mu_g^u}{2}=
  \frac12\left(\mu_3+\mu_e\right)\;.
\end{eqnarray}
\end{subequations}
Here we used Eq.\ (\ref{chemischePotentiale}).
Labelling the components of the full quark propagator with a color and flavor index is slightly misleading. For instance, red-up and green-down quasiparticles are both admixtures of red up and green down quarks. Both quasiparticles have the same Fermi surface $\bar\mu$, and only their dispersion relations $k_0=-\epsilon_\fettu{k}^e\left(\bar\mu,\phi_{1c}^{\phantom{1}e}\right)\pm\delta\mu_1$ differ by $2\,\delta\mu_1$.

With Eqs.\ (\ref{Delta_n}) and (\ref{Gplus}) and the relation
$G_0^\mp \Delta_n^\pm G^\pm = G^\mp \Delta_n^\pm G_0^\pm$, Eq.\ (\ref{Xielemente}) becomes
\begin{subequations}
\label{Xiplus}
\begin{eqnarray}
  {\Xi^\pm}_{rg}^{ud}(K)&=&-\sum_{c,e}
  \frac{\mathcal{P}_{\mp c}^{\mp e}(\fettu{k})\,
  \phi_{1c}^{\phantom{1}e}(\pm K)}
  {(k_0\mp\delta\mu_1)^2
  -\left[\epsilon_\fettu{k}^e
  \left(\bar\mu,
  \phi_{1c}^{\phantom{1}e}\right)\right]^2}\;,\\
  {\Xi^\pm}_{rg}^{du}(K)&=&-\sum_{c,e}
  \frac{\mathcal{P}_{\mp c}^{\mp e}(\fettu{k})\,
  \phi_{2c}^{\phantom{2}e}(\pm K)}
  {(k_0\mp\delta\mu_2)^2
  -\left[\epsilon_\fettu{k}^e
  \left(\bar\mu,
  \phi_{2c}^{\phantom{2}e}\right)\right]^2}\;,\\
  {\Xi^\pm}_{gr}^{ud}(K)&=&-\sum_{c,e}
  \frac{\mathcal{P}_{\mp c}^{\mp e}(\fettu{k})\,
  \phi_{2c}^{\phantom{2}e}(\pm K)}
  {(k_0\pm\delta\mu_2)^2
  -\left[\epsilon_\fettu{k}^e
  \left(\bar\mu,
  \phi_{2c}^{\phantom{2}e}\right)\right]^2}\;,\\
  {\Xi^\pm}_{gr}^{du}(K)&=&-\sum_{c,e}
  \frac{\mathcal{P}_{\mp c}^{\mp e}(\fettu{k})\,
  \phi_{1c}^{\phantom{1}e}(\pm K)}
  {(k_0\pm\delta\mu_1)^2
  -\left[\epsilon_\fettu{k}^e
  \left(\bar\mu,
  \phi_{1c}^{\phantom{1}e}\right)\right]^2}\;.
\end{eqnarray}
\end{subequations}
Here we assumed that $\phi_{nc}^{\phantom{n}e}(-K)
={\phi_{nc}^{\phantom{n}e}}^*(K)$.
Taking the gluon interaction to be point-like, $D_{\mu\nu}(K-Q)=-\,g_{\mu\nu}/\Lambda^2$, $\Lambda=const.$, the Dirac structure of the gap equations (\ref{Gapkombi}) can be projected out to yield gap equations for the gap functions
$\phi_{nc}^{\phantom{n}e}(K)$. It turns out that the gap equations for different energy, $e=\pm$, and chirality, $c=\pm$, projections decouple and have the same form. We therefore omit the indices $e,c$ in the following. Due to our assumption of a point-like gluon interaction, the gap function is also independent of $K$. The gap equations (\ref{Gapkombi}) assume the simple form
\begin{subequations}
\label{Gapkombi2}
\begin{eqnarray}
  \phi_1+3\,\phi_2&=&\frac83\,\frac{g^2}{\Lambda^2}\,\frac{T}{V}
  \sum_Q\sum_e\frac{\phi_1}{(q_0-\delta\mu_1)^2
  -\left[\epsilon_\fettu{q}^e(\bar\mu,\phi_1)\right]^2}\;,\\
  3\,\phi_1+\phi_2&=&\frac83\,\frac{g^2}{\Lambda^2}\,\frac{T}{V}
  \sum_Q\sum_e\frac{\phi_2}{(q_0+\delta\mu_2)^2
  -\left[\epsilon_\fettu{q}^e(\bar\mu,\phi_2)\right]^2}\;.
\end{eqnarray}
\end{subequations}
We now perform the Matsubara sums with the help of the relation
\begin{equation}
  \label{residuum}
  T\sum_n\frac{1}{(q_0-\delta\mu)^2-\epsilon^2}=
  -\,\frac{1}{4\epsilon}
  \left[\tanh\left(\frac{\epsilon+\delta\mu}{2T}\right)
  +\tanh\left(\frac{\epsilon-\delta\mu}{2T}\right)\right]
  \newatop{\displaystyle\longrightarrow}{T\rightarrow 0}{-1}
  -\frac{1}{2\epsilon}\,
  \theta(\epsilon-|\delta\mu|)\;.
\end{equation}
The resulting gap equations are
\begin{subequations}
\label{Gaps}
\begin{eqnarray}
  \phi_1+3\,\phi_2&=&-\,\frac{2}{3\pi^2}\,
  \frac{g^2}{\Lambda^2}\sum_e
  \int_0^\kappa\ud q\,q^2\,\frac{\phi_1}{\epsilon_\fettu{q}^e
  (\bar\mu,\phi_1)}\,\theta\left(\epsilon_\fettu{q}^e
  (\bar\mu,\phi_1)-|\delta\mu_1|\right)\;,\\
  3\,\phi_1+\phi_2&=&-\,\frac{2}{3\pi^2}\,
  \frac{g^2}{\Lambda^2}\sum_e
  \int_0^\kappa\ud q\,q^2\,\frac{\phi_2}{\epsilon_\fettu{q}^e
  (\bar\mu,\phi_2)}\,\theta\left(\epsilon_\fettu{q}^e
  (\bar\mu,\phi_2)-|\delta\mu_2|\right)\;.
\end{eqnarray}
\end{subequations}
Here we introduced a cutoff $\kappa$ to render the momentum integral finite. 

The equation of state for 2SC matter, i.e., the pressure as a function of temperature and chemical potential, results from the relation
\begin{equation}
  p_\mathrm{2SC}=\frac{T}{V}\,\Gamma^*\;,
\end{equation}
where $\Gamma^*$ is the value of the effective action (\ref{Gamma}) at the stationary point determined by Eqs.\ 
(\ref{stationary}).
Since we do not consider the gluons as dynamical degrees of freedom, the first three terms in Eq.\ (\ref{Gamma}) can be omitted. The last two terms in Eq.\ (\ref{Gamma}) can be simplified with the help of the Dyson-Schwinger equation (\ref{DSE}). The final result reads
\begin{equation}
  \label{p}
  p_\mathrm{2SC}=\frac12\,\frac{T}{V}\left[\Sp\ln S^{-1}
  -\frac12\,\Sp\,(\Sigma\,S)\right]\;,
\end{equation}
where the propagator $S$ obeys the Dyson-Schwinger equation (\ref{DSE}). Performing the trace over Nambu-Gor'kov space, the second term can be written as
\begin{equation}
  \label{SigmaS}
  -\,\frac14\,\frac{T}{V}\,    
  \Sp\,(\Sigma\,S)=-\,\frac14\,\frac{T}{V}\,
  \newatop{\displaystyle\Sp}{X,c,f,s}{-1}\left(
  \Sigma^+\,G^+ + \Sigma^-\, G^- + \Phi^- \, \Xi^+ + \Phi^+ \, 
  \Xi^- \right)\;,
\end{equation}
where the trace on the right-hand side runs over space-time, color, flavor, and Dirac indices. Since we neglected the regular self-energies $\Sigma^\pm$ in the derivation of the gap equation, to be consistent we also have to drop the first two terms in Eq.\ (\ref{SigmaS}). Fourier-transforming into momentum space and using the gap equations (\ref{Gapgleichung}), we obtain
\begin{equation}
  -\,\frac14\,\frac{T}{V}\,
  \Sp\,(\Sigma\,S)
  =-\,\frac{g^2}{4}\,\frac{T^2}{V^2}\sum_{K,Q}
  \newatop{\displaystyle\sp}{c,f,s}{-1}
  \left[\gamma^\mu T_a\,\Xi^-(K)\,\gamma^\nu
   T_b^T\, \Xi^+(Q)+
  \gamma^\mu T_a^T\,\Xi^+(K)\,\gamma^\nu T_b\, 
  \Xi^-(Q)\right]\,D_{\mu\nu}^{ab}(K-Q)\;,
\end{equation}
where the trace on the right-hand side runs only over color, flavor, and Dirac indices. We now insert the local, instantaneous gluon propagator $D_{\mu\nu}^{ab}(K-Q)=-\,\delta^{ab}g_{\mu\nu}/\Lambda^2$, and sum over $a,b$ and $\mu,\nu$. 
Then we perform the trace over color, flavor and Dirac space with the help of Eqs.\ (\ref{Ximatrix}), (\ref{Xiplus}). Due to the point-like gluon interaction the sums over $K$ und $Q$ separate. These sums can be simplified with the help of the gap equations (\ref{Gapkombi2}). The final result is
\begin{equation}
\label{p2}
-\,\frac14\,\frac{T}{V}\,\Sp\,(\Sigma\,S)=\frac34\,\frac{\Lambda^2}{g^2}\,\left[|\phi_1|^2+|\phi_2|^2+3\,(\phi_1^*\phi_2+\phi_1\phi_2^*)\right]\;.
\end{equation}
The first term in Eq.\ (\ref{p}) is straightforwardly evaluated as
\begin{eqnarray}
\lefteqn{\frac12\,\frac{T}{V}\,\Sp\ln S^{-1}=}\nonumber\\
&=&\frac{T}{V}\sum_K\sum_{e}\left\{\sum_{n=1,2}\sum_{j=\pm}
\ln\left[\frac{(k_0+j\,\delta\mu_n)^2-
[\epsilon_\fettu{k}^e(\bar\mu,\phi_n)]^2}{T^2}\right]
+\ln\left[\frac{k_0^2-
[\epsilon_\fettu{k}^e(\mu_b^u,0)]^2}{T^2}\right]
+\ln\left[\frac{k_0^2-
[\epsilon_\fettu{k}^e(\mu_b^d,0)]^2}{T^2}\right]
\right\}\nonumber\\
&=&\frac{1}{\pi^2}\sum_e\int_0^\kappa\ud k\,k^2 \left( \sum_{n=1,2} \left\{ 
\epsilon_\fettu{k}^e(\bar\mu,\phi_n)+\sum_{j=\pm} T\,\ln\left[1+ \exp \left(-\,\frac{\epsilon_\fettu{k}^e(\bar\mu,\phi_n)+j\,\delta\mu_n}{T}\right)\right] \right\}\right.\nonumber\\
&&\hspace*{5.34cm}
 +\, \epsilon_\fettu{k}^e(\mu_b^u,0) + 2\,T\,\ln\left[1+ \exp \left(-\,\frac{\epsilon_\fettu{k}^e(\mu_b^u,0)}{T}\right)\right]
\nonumber\\
&&\hspace*{5.34cm}
+\left.\epsilon_\fettu{k}^e(\mu_b^d,0) + 2\,T\,\ln\left[1+ \exp \left(-\,\frac{\epsilon_\fettu{k}^e(\mu_b^d,0)}{T}\right)\right]
\right)\;.
\label{p1}
\end{eqnarray}
Subtracting the contribution from the vacuum and taking the limit $T\rightarrow 0$ leads to
\begin{equation}
  \frac12\,\frac{T}{V}\,\Sp\ln S^{-1}=
  \frac{1}{\pi^2}\sum_e\sum_{n=1,2}\int_0^\kappa\ud k\,k^2
  \left[\epsilon_\fettu{k}^e(\bar\mu,\phi_n)-k+\frac{k}{3}\,
  \frac{k-e\bar\mu}{\epsilon_\fettu{k}^e(\bar\mu,\phi_n)}\,
  \theta\left(|\delta\mu_n|-
  \epsilon_\fettu{k}^e(\bar\mu,\phi_n)\right)\right]
  +\frac{{\mu_b^u}^4+{\mu_b^d}^4}{12\pi^2}\; .
\end{equation}
To obtain the pressure (\ref{p}) for color-superconducting quark matter with two flavors we have to add Eqs.\ (\ref{p2}) and (\ref{p1}).

In Sec.\ \ref{results} we shall consider compact stellar objects which have to be neutral with respect to electric charge. In order to achieve this, we have to add the contribution of electrons to the pressure (\ref{p}) of our two-flavor color superconductor,
\begin{equation}
  \label{pe}
  p_e=\frac{\mu_e^4}{12\pi^2}\; ,
\end{equation}
where we neglected the small electron mass. If the chemical potential for strange quarks, $\mu_i^s=\mu+\mu_e/3+\mu_3 T_{ii}^3+\mu_8 T_{ii}^8$, exceeds the strange quark mass, $m_s$, we also have to include strange quarks into our consideration. We assume them to be non-interacting, which leads to the following additional contribution to Eq.\ (\ref{p}),
\begin{equation}
\label{ps}
p_s=\frac{1}{3\pi^2}\sum_{i\,=\,r}^b\int_0^{{k_F}_i^s}\ud k\,\frac{k^4}{E_\fettu{k}^s}\; ,
\end{equation}
with the Fermi momentum
${k_F}_i^s=({\mu_i^s}^2-m_s^2)^{1/2}$
and the energy $E_\fettu{k}^s=(k^2+m_s^2)^{1/2}$.
Strange quarks also serve to neutralize the large positive electric charge of a system of up and down quarks. Consequently, we expect the electron density to be reduced once strange quarks are present in the system.
The total pressure of our system is the sum of Eqs.\ (\ref{p}), (\ref{pe}), and (\ref{ps}),
\begin{equation}
\label{totalp}
  p=p_\mathrm{2SC}+p_e+p_s-B\; .
\end{equation}
Here, we have also subtracted the pressure of the perturbative vacuum in the form of the MIT bag constant $B$ \cite{MIT}. This will prove essential to obtain bound stars of finite radius.

Compact stellar objects are not only neutral with respect to electric charge but also with respect to color charge. The neutrality conditions read
\begin{subequations}
\begin{eqnarray}
  \label{eneut}
  n_e&\equiv&\frac{\partial p}{\partial\mu_e}=0\; ,\\
  \label{3neut}
  n_3&\equiv&\frac{\partial p}{\partial\mu_3}=0\; ,\\
  \label{8neut}
  n_8&\equiv&\frac{\partial p}{\partial\mu_8}=0\; ,
\end{eqnarray}
\end{subequations}
where $n_e$ is the total electric charge density and $n_3$, $n_8$ are the color charge densities.
It is straightforward to see that the solution of Eq.\ (\ref{3neut}) is $\mu_3=0$. The reason is that in a two-flavor color superconductor, $SU(3)_c$ is broken to $SU(2)_c$. One of the generators of this residual $SU(2)_c$ symmetry is $T_3$ and, consequently, the associated color-chemical potential has to vanish (otherwise, $SU(2)_c$ would be broken explicitly). We therefore have $\mu_3=0$ irrespective of whether we enforce color neutrality or not.
>From Eqs.\ (\ref{muquer}) we then conclude $\delta\mu\equiv\delta\mu_1=-\delta\mu_2=-\mu_e/2$. Inserting this result into the gap equations (\ref{Gaps}) we read off that the only possible solution is $\phi\equiv\phi_1=-\phi_2$.
This greatly simplifies the gap equations; there is only a single gap equation for $\phi$,
\begin{equation}
\label{finalGap}
3\pi^2\frac{\Lambda^2}{g^2}=\sum_e\int_0^\kappa \ud
  q\,q^2\,\frac{1}{\epsilon_\fettu{q}^e
  (\bar\mu,\phi)}\,\theta\left(\epsilon_\fettu{q}^e
  (\bar\mu,\phi)-|\delta\mu|\right)\;,
\end{equation}
and the expression (\ref{p2}) becomes
\begin{equation}
-\,\frac14\,\frac{T}{V}\,\Sp\,(\Sigma\,S)=-\,3\,
\frac{\Lambda^2}{g^2}\,|\phi|^2\;.
\end{equation}
Finally, in order to solve the Tolman-Oppenheimer-Volkoff equation, we need the energy density, which at $T=0$ reads
\begin{equation}
\label{edichte}
  \varepsilon=\mu n+\mu_3 n_3+\mu_8 n_8+\mu_e n_e-p\;.  
\end{equation}
Here, $n\equiv\partial p/\partial\mu$ is the quark density.
\section{Results}
\label{results}
In this section we numerically solve the gap equation (\ref{finalGap}) and compute the equation of state (\ref{totalp}). With Eqs.\ (\ref{totalp}) and (\ref{edichte}) we then solve the TOV equation to obtain the mass-radius relation for quark stars. Unless mentioned otherwise, in order to compare our results to those of Ref.\ \cite{MeiHuang},
we use the following values for the parameters of our model,
\begin{subequations}
\begin{eqnarray}
  m_s&=&0.1407\ \mathrm{GeV}\; ,\\
  \label{coupling}
  g^2/\Lambda^2&=&45.1467\ \mathrm{GeV}^{-2}\; ,\\
  \kappa&=&0.6533\ \mathrm{GeV}\; ,\\
  B^{1/4}&=&0.17\ \mathrm{GeV}\; .
\end{eqnarray}
\end{subequations}

\begin{figure}[!hbp]
  \begin{center}
    \includegraphics[scale=1.0]{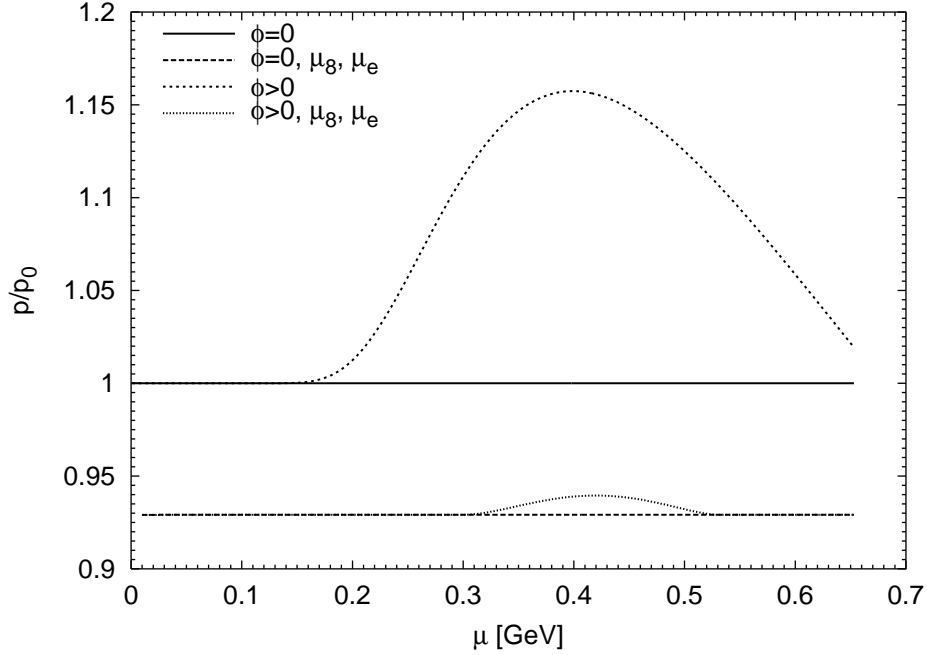}
    \caption{The pressure as a function of the quark-chemical    potential, normalized to its value for vanishing gap and $\mu_e=\mu_8=0$. The contribution of strange quarks is omitted. The bag constant $B$ is set to zero. Full and dashed lines are obtained for normal-conducting quark matter, $\phi=0$. Short-dashed and dotted lines are for color-superconducting quark matter, $\phi>0$. Full lines and short-dashed lines are computed without the constraints of electric- and color-charge neutrality, $\mu_e=\mu_8=0$. Dashed and dotted lines are computed for an electric- and color-charge neutral system, $\mu_e\neq 0$, $\mu_8\neq 0$.}
    \label{pdpref2mu}
  \end{center}
\end{figure}

\begin{figure}[!hbp]
  \begin{center}
    \includegraphics[scale=1.0]{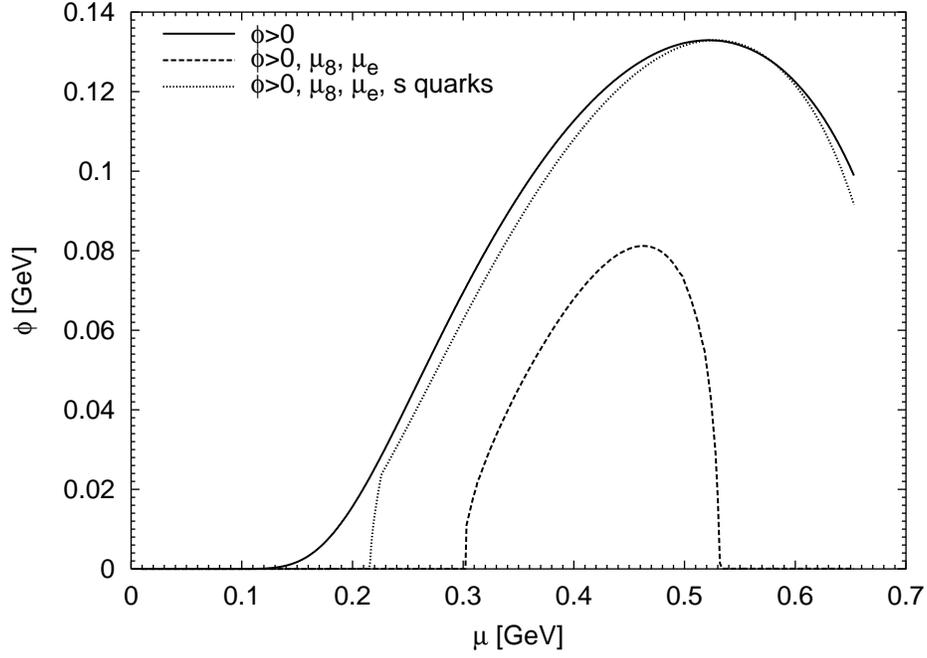}
    \caption{The solution $\phi$ of the gap equation (\ref{finalGap}) as a function of the quark-chemical potential $\mu$. The full line is for $\mu_e=\mu_8=0$. For the dashed and dotted lines $\mu_e$ and $\mu_8$ are determined from the conditions (\ref{eneut}) and (\ref{8neut}) of electric- and color-charge neutrality. The difference between dashed and dotted lines is that for the former the contribution of strange quarks is neglected, while for the latter it is included.}
    \label{phi2mu}
  \end{center}
\end{figure}

In Fig.\ \ref{pdpref2mu} we show the total pressure (\ref{totalp}), normalized to its value for $\phi=\mu_e=\mu_8=0$, as a function of the quark-chemical potential for various cases. The contribution of strange quarks is omitted, $p_s=0$, and the bag constant is set to zero, $B=0$. One observes that the color-superconducting state with $\phi>0$ has a larger pressure than the normal-conducting state with $\phi=0$. Consequently, the color-superconducting state is energetically preferred. The difference in pressure between color-superconducting and normal-conducting states is proportional to the value of the gap (squared), cf.\ Fig.\ \ref{phi2mu}. The constraint of electric-charge neutrality reduces the pressure. (The effect on the pressure when imposing color-charge neutrality is negligibly small.) This constraint is necessary to obtain stable stars. Electrically charged stars would explode because of the repulsive Coulomb force. (One also has to impose color-charge neutrality because color-charged stars cannot exist due to confinement.)

\begin{figure}[!hbp]
  \begin{center}
    \includegraphics[scale=1.0]{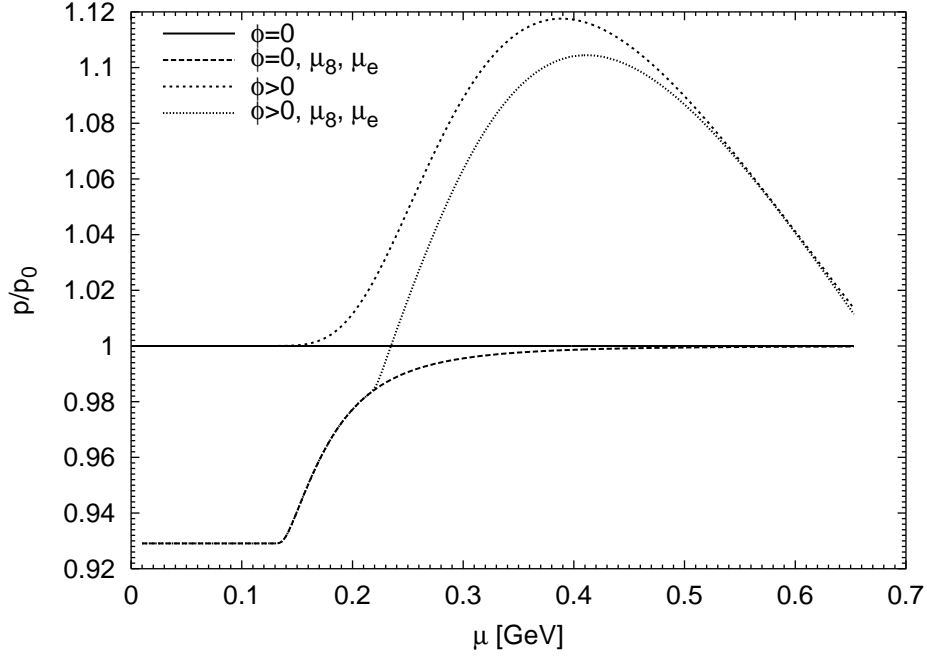}
    \caption{The same as in Fig.\ \ref{pdpref2mu}, but with the contribution of strange quarks.}
    \label{pdpref2mumits}
  \end{center}
\end{figure}

\begin{figure}[!hbp]
  \begin{center}
    \includegraphics[scale=1.0]{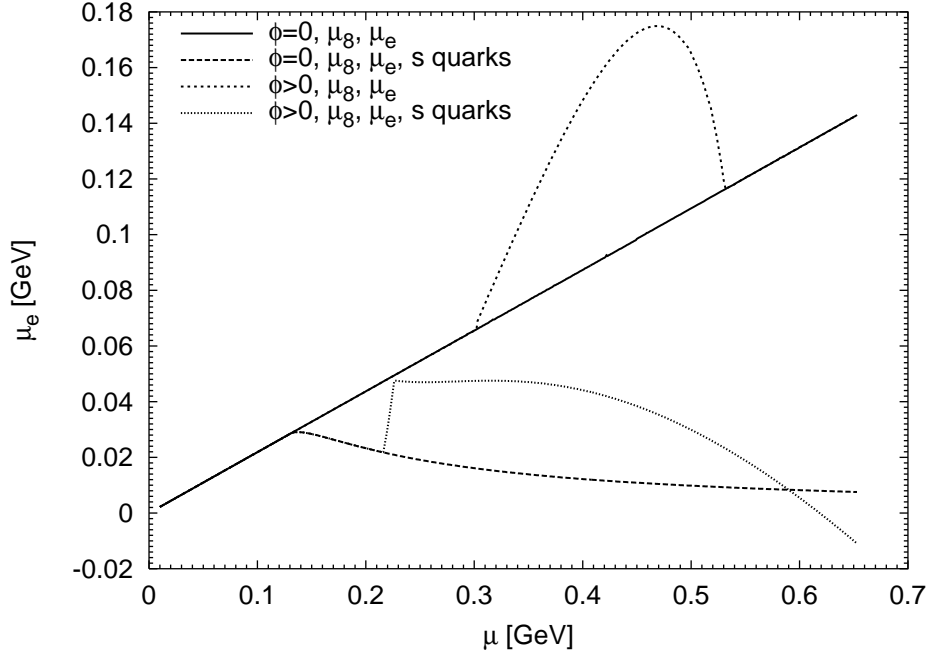}
    \caption{The electro-chemical potential $\mu_e$ as a function of the quark-chemical potential $\mu$ when enforcing the conditions of electric- and color-charge neutrality (\ref{eneut}) and (\ref{8neut}). The full line and the dashed line are for normal-conducting quark matter, while the short-dashed and dotted line are for color-superconducting quark matter. The full line and the short-dashed line are obtained neglecting strange quarks, the dashed and dotted line are obtained including strange quarks.}
    \label{mue2mu}
  \end{center}
\end{figure}

In Fig.\ \ref{pdpref2mumits} the same cases are shown as in Fig.\ \ref{pdpref2mu}, now including the contribution of strange quarks. Again, the color-superconducting state is energetically preferred over the normal-conducting state. The difference to the previous case is that, as soon as $\mu$ exceeds $m_s$, strange quarks partially assume the role of electrons to ensure electric-charge neutrality of the system. Consequently, the electronic contribution to the pressure is reduced, cf.\ Fig.\ \ref{mue2mu}, and the reduction of the pressure when imposing electric-charge neutrality becomes smaller. For large values of the quark-chemical potential, $\mu\gg m_s$, the amount of electrons necessary to make the system electrically neutral becomes negligibly small. Therefore, the full and dashed lines, as well as the short-dashed and dotted lines in Fig.\ \ref{pdpref2mumits} approach each other.

In Fig.\ \ref{phi2mu} we show the gap as a function of the quark-chemical potential. For small values of the quark-chemical potential the gap vanishes. At larger values of $\mu$, the gap increases, until the quark-chemical potential approaches the value of the cutoff $\kappa$, where due to restricted phase space the gap starts to decrease. The conditions of electric and color neutrality (\ref{eneut}) and (\ref{8neut}) tend to decrease the value of the gap. This decrease is especially pronounced in the case without strange quarks (dashed line), where the electro-chemical potential is large, see Fig.\ \ref{mue2mu}. Including strange quarks, the electro-chemical potential becomes smaller and, consequently, the gap becomes larger (dotted line). Note the kink in the dotted line in Fig.\ \ref{phi2mu}. To the left of the kink, the system is in the so-called ''gapless'' color-superconducting phase discussed in detail in Ref.\ \cite{Igor}, while to the right it is in the standard color-superconducting phase without gapless modes. For the dashed line, the system is always in the gapless color-superconducting phase.

Figure \ref{mue2mu} shows the value of the electro-chemical potential $\mu_e$ as a function of $\mu$ as obtained from enforcing the neutrality conditions (\ref{eneut}) and (\ref{8neut}). For normal-conducting quark matter without strange quarks, electric-charge neutrality requires that the electron density increases proportional to the quark density. Consequently, $\mu_e$ is a linearly rising function of $\mu$ (full line). In the presence of strange quarks the amount of electrons required to achieve electric neutrality is smaller. Therefore, $\mu_e$ decreases as soon as $\mu_i^s>m_s$. For an electric- and color-charge neutral color superconductor either with or without strange quarks, the electro-chemical potential increases substantially as compared to a normal conductor. (Only with strange quarks and for very large values of $\mu$ close to $\kappa$, the value of $\mu_e$ in the color-superconducting phase is smaller than in a normal conductor and may even become negative. We perceive the latter to be an artefact of approaching the limit of phase space.)

\begin{figure}[!hbp]
  \begin{center}
    \includegraphics[scale=1.0]{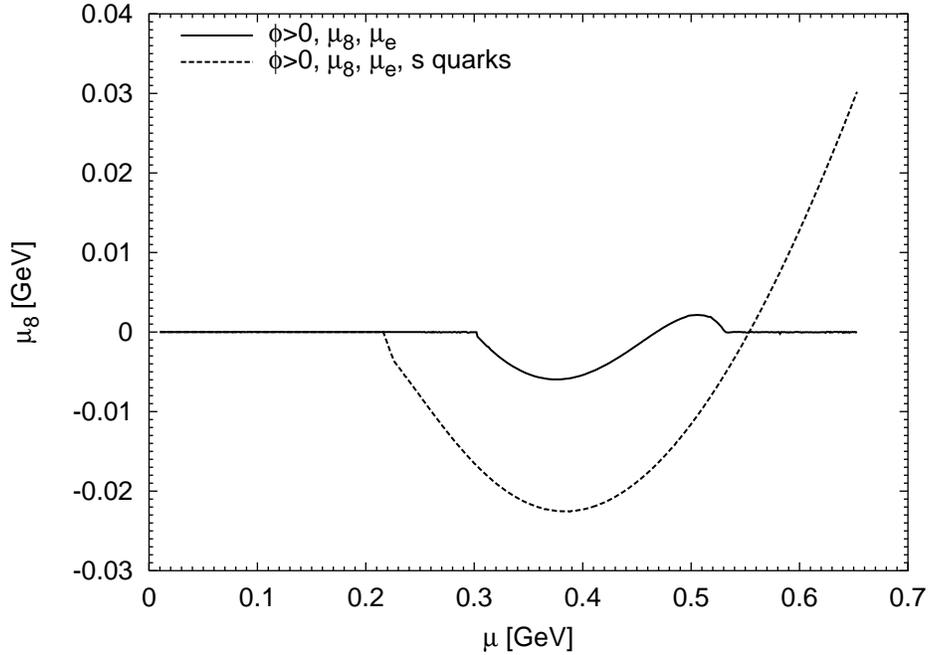}
    \caption{The color-chemical potential $\mu_8$ as a function of the quark-chemical potential $\mu$, for electric- and color-neutral color-superconducting quark matter without strange quarks (full line) and with strange quarks (dashed line).}
    \label{mu82mu}
  \end{center}
\end{figure}

In Fig.\ \ref{mu82mu} we show the value of the color-chemical potential $\mu_8$ as a function of $\mu$ for electric- and color-neutral color-superconducting quark matter with and without strange quarks. (Normal-conducting matter is automatically color neutral in the thermodynamic limit.) The values of $\mu_8$ necessary to make the system color neutral are much smaller than the values of $\mu_e$ required for electric neutrality. This can be understood in the weak-coupling limit where the gap is much smaller than the chemical potential, $\phi\sim\mu\,\exp(-1/g^2)\ll\mu$. In this limit, the value of $\mu_8$ required to achieve color neutrality is parametrically of order $\mu_8\sim\phi^2/\mu$ \cite{Rebhan}. 

\begin{figure}[!hbp]
  \begin{center}
    \includegraphics[scale=1.0]{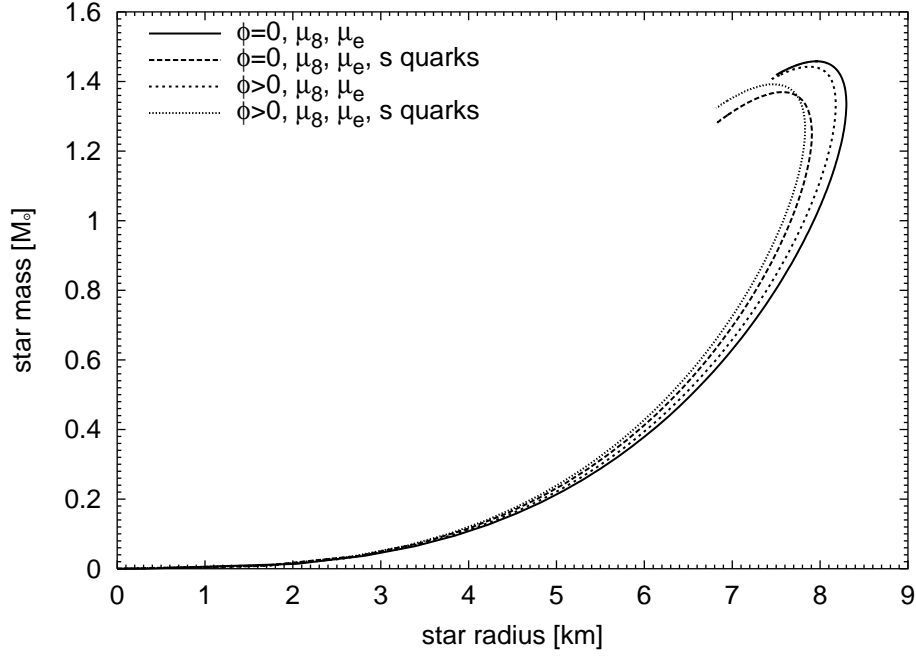}
    \caption{The mass-radius relation for electric- and color-neutral quark stars. Full and dashed lines are for normal-conducting quark matter, short-dashed and dotted lines are for color-superconducting quark matter. Full and short-dashed lines are computed without, dashed and dotted lines are computed with strange quarks.}
    \label{M2R}
  \end{center}
\end{figure}

With Eqs.\ (\ref{totalp}) and (\ref{edichte}) we now solve the TOV equations
\begin{subequations}
\begin{eqnarray}
  \label{TOV-Gleichung}
  \frac{\ud p}{\ud r}&=&-\,\frac{[p(r)+\varepsilon(r)]
  [M(r)+4\pi r^3p(r)]}{r[r-2M(r)]}\;,\\
  M(r)&=&4\pi\int_0^r
  \varepsilon(r^\prime){r^\prime}^2\,\ud r^\prime\;.
\end{eqnarray}
\end{subequations}
The resulting mass-radius relations are shown in Fig.\ \ref{M2R}. We observe that the influence of color superconductivity on the mass-radius relation is at most on the order of a few percent, in agreement with the results of Ref.\ \cite{Blaschke}. This was to be expected, since superconductivity is a Fermi-surface phenomenon, while the equation of state which determines the mass and radius is sensitive to the whole Fermi sea. To be more precise, the relative change in the pressure due to superconductivity is of the order $\phi^2/\mu^2$. For $\phi\ll\mu$ this is a tiny effect. Indeed, as can be seen from Fig.\ \ref{M2R}, including strange quarks has a much larger effect on the masses and radii.
The reason is that adding strange quarks to the system affects a relative change of the pressure by a term $\sim(\mu_s/\mu)^4\sim\mathcal{O}(1)$.

The observation that color superconductivity does not have an effect
on the mass-radius relation of a quark star for $\phi\ll\mu$
immediately leads to the question how large the gap has to become in
order to see an appreciable change in either the mass or the radius of
the star. To our knowledge this question has so far not been addressed
in the literature and is the main motivation for our current
study. This question can be answered by artificially increasing the
coupling constant $g^2/\Lambda^2$, such that the solution $\phi$ of
the gap equation (\ref{finalGap}) increases as well. 
Of course, such a modification  will
change the vacuum properties, such as the pion decay constant, 
within our NJL model. However, at nonzero baryon density, this model is 
in any case not very realistic as it neither describes saturation of 
nuclear matter in the ground state nor does it feature a 
hadronic phase where chiral symmetry is broken.
Here, we consider the NJL model just as an effective model to describe
quark-quark interactions at high baryon density. 
It is then certainly a permissible and interesting question to ask
how the coupling strength affects the value of the
color-superconducting gap parameter and quark star properties.

\begin{figure}[!hbp]
  \begin{center}
    \includegraphics[scale=1.0]{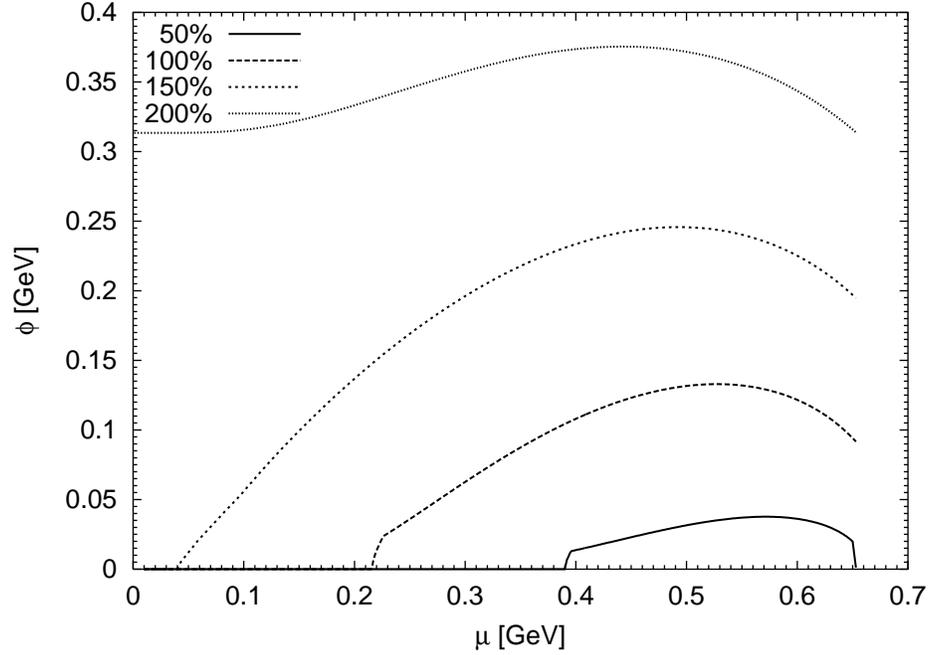}
    \caption{The gap $\phi$ as a function of the quark-chemical potential $\mu$. Full line: the coupling constant $g^2/\Lambda^2$ is reduced by a factor of two as compared to the default value (\ref{coupling}), for which the corresponding gap is shown as the dashed line. Short-dashed line: the coupling constant is 3/2 times the default value; dotted line: the coupling constant is twice the default value. All gaps are for electric- and color-neutral matter including strange quarks.}
    \label{phi2muprozent}
  \end{center}
\end{figure}

In Fig.\ \ref{phi2muprozent} we show the solution of the gap equation when decreasing the coupling constant by a factor of two (full line) as well as increasing it by a factor of $3/2$ (short-dashed line) and a factor of two (dotted line), respectively.
In the latter case we observe the interesting phenomenon that the gap is non-vanishing even when $\mu=0$. At first sight this seems surprising but one can readily convince oneself that the gap equation (\ref{finalGap}) indeed has a non-trivial solution for $\mu=0$, if the coupling constant exceeds the critical value $g^2/\Lambda^2=3\pi^2/\kappa^2\simeq 69.37$ GeV$^{-2}$. The situation where the gap was non-vanishing in the vacuum was also analyzed in Ref.\ \cite{CarterandDiakonov}.

\begin{figure}[!hbp]
  \begin{center}
    \includegraphics[scale=1.0]{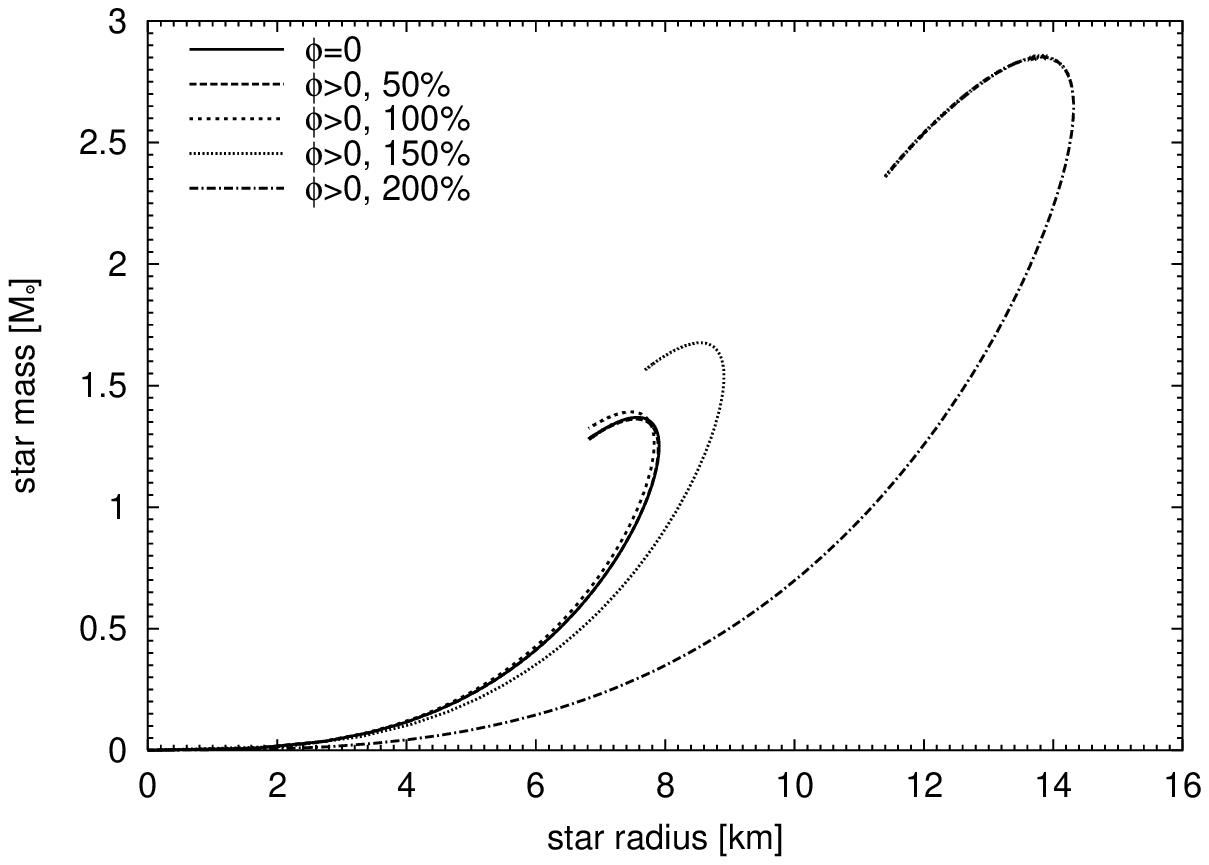}
    \caption{The mass-radius relation of electric- and color-neutral
quark stars containing strange quarks. The full line is for
normal-conducting quark stars, all other lines are for
color-superconducting quark stars. Dashed line (almost indistinguishable from the full line): the coupling constant is 50\% of its default value. Short-dashed line: the coupling constant assumes its default value. Dotted line: the coupling constant is multiplied by a factor 3/2. Dash-dotted line: the coupling constant is twice as large as the default value.}
    \label{M2Rprozent}
  \end{center}
\end{figure}

In Fig.\ \ref{M2Rprozent} we show the mass-radius relations calculated with the equations of state corresponding to the gaps shown in Fig.\ \ref{phi2muprozent}. As expected, a decrease of the coupling constant by a factor of two does not appreciably change the mass-radius relation, but multiplying the coupling constant by a factor 3/2 already leads to an increase of maximum mass and radius by $\sim 20$\%. For the coupling constant which is a factor of two larger than the default value we find that the quark star also doubles in mass and radius. The reason is that the equation of state changes appreciably for gaps of order $\sim 300$ MeV, cf.\ Fig.\ \ref{phi2muprozent}, since then $\phi\sim\mu$.
\section{Conclusions}
\label{conclusions}
In this paper we have investigated color-superconducting quark matter in the so-called 2SC phase, i.e., a color superconductor consisting of massless up and down quarks, where quarks of red and green color form anti-blue Cooper pairs. Starting from the CJT formalism, we have derived the gap equation and the pressure. By adding electrons and/or strange quarks, we have also studied this system under the constraints of electric and color neutrality. For electric- and color-neutral systems, we have solved the TOV equation in order to determine the mass-radius relation for quark stars. We confirmed the result of Ref.\ \cite{Blaschke} that color superconductivity does not substantially alter the mass and the radius of a quark star, if the coupling constant is chosen to reproduce vacuum properties such as the pion decay constant. The reason is that superconductivity has an effect of the equation of state which is proportional to $\phi^2/\mu^2$ which is small if $\phi\ll\mu$.

We then asked the question how large the color-superconducting gap parameter has to be in order to see an appreciable effect on the mass and the radius of a quark star. To this end, we artificially increased the coupling constant which has the effect of increasing the color-superconducting gap parameter. We found non-trivial solutions of the gap equation in the vacuum, i.e., for $\mu=0$, when the coupling constant exceeds a critical value which is directly proportional to the cutoff parameter of our NJL-type model. For gaps of the order of 300 MeV the mass and radius of a color-superconducting quark star was found to be twice as large as for a normal-conducting quark star. While this is \emph{per se} an interesting result, such quark stars are still of the same size and mass as ordinary neutron stars. It is thus impossible to decide whether a compact stellar object consists of normal-conducting or color-superconducting quark matter, or simply of ordinary neutron matter.
\section*{Acknowledgments}
The authors thank Debades Bandyopadhyay, Matthias Hanauske, Mei Huang, J\"urgen Schaffner-Bielich, and Igor Shovkovy
for discussions.




\begin{thebibliography}{99}

\bibitem{BailinLove}D.\ Bailin, A.\ Love, 
Phys.\ Rept.\ \textbf{107}, 325 (1984)

\bibitem{Prakash}D.\ Page, M.\ Prakash, J.M.\ Lattimer, A.\ Steiner, Phys.\ Rev.\ Lett.\ \textbf{85}, 2048 (2000)

\bibitem{Steiner}A.W.\ Steiner, S.\ Reddy, M.\ Prakash,
Phys.\ Rev.\ D \textbf{66}, 094007 (2002)

\bibitem{buballaoertel}M.\ Buballa, M.\ Oertel,
Nucl.\ Phys.\ \textbf{A 703}, 770 (2002)

\bibitem{Baldo}M.\ Baldo, M.\ Buballa, F.\ Burgio, F.\ Neumann, M.\ Oertel, H.J.\ Schulze, Phys.\ Lett.\ \textbf{B 562}, 153 (2003)

\bibitem{LugonesHorvath1}J.E.\ Horvath, G.\ Lugones, J.A.\ de Freitas Pacheco, Int.\ J.\ Mod.\ Phys.\ \textbf{D 12}, 519 (2003)

\bibitem{LugonesHorvath2}G.\ Lugones, J.E.\ Horvath, astro-ph/0211638

\bibitem{AlfordReddy}M.\ Alford, S.\ Reddy, Phys.\ Rev.\ D \textbf{67}, 074024 (2003)

\bibitem{Grigorian1}H.\ Grigorian, D.\ Blaschke, D.N.\ Aguilera, astro-ph/0303518

\bibitem{Banik}S.\ Banik, D.\ Bandyopadhyay, astro-ph/0212340

\bibitem{Grigorian2}
D.\ Blaschke, H.\ Grigorian, D.N.\ Aguilera, S.\ Yasui, H.\ Toki,
AIP Conf.\ Proc.\ \textbf{660}, 209 (2003)

\bibitem{Blaschke}D.\ Blaschke, S.\ Fredriksson, H.\ Grigorian, A.\ M.\ \"Ozta\c s, nucl-th/0301002

\bibitem{Shovkovy}I.\ Shovkovy, M.\ Hanauske, M.\ Huang,
Phys. Rev. D \textbf{67}, 103004 (2003)

\bibitem{CFL}M.\ Alford, K.\ Rajagopal, F.\ Wilczek,
Nucl.\ Phys.\ \textbf{B 537}, 443 (1999)

\bibitem{NJL}Y.\ Nambu, G.\ Jona-Lasinio, Phys.\ Rev.\ \textbf{122}, 345 (1961); Phys.\ Rev.\ \textbf{124}, 246 (1961)

\bibitem{RappSchaeferShuryakVelkovsky}R.\ Rapp, T.\ Sch\"afer, E.V.\ Shuryak, M.\ Velkovsky, Phys.\ Rev.\ Lett.\ \textbf{81}, 53 (1998); M.\ Alford, K.\ Rajagopal, F.\ Wilczek, Phys.\ Lett.\ \textbf{B 422}, 247 (1998)

\bibitem{Absence}M.\ Alford, K.\ Rajagopal, JHEP \textbf{0206}, 031 (2002)

\bibitem{CJT}J.M.\ Cornwall, R.\ Jackiw, E.\ Tomboulis, Phys.\ Rev.\ D \textbf{10}, 2428 (1974)

\bibitem{RischkeReview}D.H.\ Rischke, nucl-th/0305030

\bibitem{MiranskiShovkovy}V.A.\ Miransky, I.A.\ Shovkovy, L.C.R.\ Wijewardhana, Phys.\ Rev.\ D \textbf{64}, 096002 (2001)

\bibitem{Tagaki}S.\ Takagi, Prog.\ Theor.\ Phys.\ \textbf{109}, 233 (2003)

\bibitem{Japaner}H.\ Abuki, hep-ph/0306074 

\bibitem{Rebhan}A.\ Gerhold and A.\ Rebhan, hep-ph/0305108;
A. Kryjevski, hep-ph/0305173;
D.D.\ Dietrich and D.H.\ Rischke (in preparation)

\bibitem{Manuel}C.\ Manuel, Phys.\ Rev.\ D \textbf{62}, 114008 (2000)

\bibitem{Rockefeller}W.E.\ Brown, J.T.\ Liu, H.\ Ren, Phys.\ Rev.\ D \textbf{61}, 114012 (2000); Phys.\ Rev.\ D \textbf{62}, 054013 (2000); Phys.\ Rev.\ D \textbf{62}, 054016 (2000)

\bibitem{Wang}Q.\ Wang, D.H.\ Rischke, Phys.\ Rev.\ D \textbf{65}, 054005 (2002)

\bibitem{Rischkeselfenergy}D.H.\ Rischke, Phys.\ Rev.\ D \textbf{64}, 094003 (2001)

\bibitem{MeiHuang}M.\ Huang, P.\ Zhuang, W.\ Chao, Phys.\ Rev.\ D \textbf{67}, 065015 (2003)

\bibitem{Igor}I.\ Shovkovy, M.\ Huang, hep-ph/0302142

\bibitem{CarterandDiakonov}G.W.\ Carter, D.\ Diakonov, Phys.\ Rev.\ D \textbf{60}, 016004 (1999)

\bibitem{Rischke-Paper}R.D.\ Pisarski, D.H.\ Rischke, Phys.\ Rev.\ Lett.\ \textbf{83}, 37 (1999); Phys.\ Rev.\ D \textbf{60}, 094013 (1999)

\bibitem{MIT}A.\ Chodos, R.L.\ Jaffe, K.\ Johnson, C.B.\ Thorn, V.F.\ Weisskopf, Phys.\ Rev.\ D \textbf{9}, 3471 (1974)

\end{thebibliography}
\end{document}